\documentclass[a4paper]{article}

\usepackage{INTERSPEECH2022}

\usepackage{cite}
\usepackage{multirow}
\usepackage{todonotes}
\usepackage[
separate-uncertainty = true,
multi-part-units = repeat
]{siunitx}
\usepackage{stfloats}
\usepackage{graphicx}
\usepackage{subfig}

\title{SOMOS: The Samsung Open MOS Dataset\\ for the Evaluation of Neural Text-to-Speech Synthesis}
\name{Georgia Maniati$^1$, 
	Alexandra Vioni$^1$,
	Nikolaos Ellinas$^1$,
	Karolos Nikitaras$^1$,
	Konstantinos Klapsas$^1$,
	June Sig Sung$^2$,
	Gunu Jho$^2$,
	Aimilios Chalamandaris$^1$,
	Pirros Tsiakoulis$^1$}
\address{
	$^1$Innoetics, Samsung Electronics, Greece\\
	$^2$Mobile Communications Business, Samsung Electronics, Republic of Korea}

\email{g.maniati@samsung.com, a.vioni@partner.samsung.com}

\begin{document}
	
	\maketitle
	\begin{abstract}
		
		In this work, we present the SOMOS dataset, the first large-scale mean opinion scores (MOS) dataset consisting of solely neural text-to-speech (TTS) samples.
		It can be employed to train automatic MOS prediction systems focused on the assessment of modern synthesizers, and can stimulate advancements in acoustic model evaluation.
		It consists of 20K synthetic utterances of the LJ Speech voice, a public domain speech dataset which is a common benchmark for building neural acoustic models and vocoders.
		Utterances are generated from 200 TTS systems including vanilla neural acoustic models as well as models which allow prosodic variations.
		An LPCNet vocoder is used for all systems, so that the samples' variation depends only on the acoustic models.
		The synthesized utterances provide balanced and adequate domain and length coverage.
		We collect MOS naturalness evaluations on 3 English Amazon Mechanical Turk locales and share practices leading to reliable crowdsourced annotations for this task.
		We provide baseline results of state-of-the-art MOS prediction models on the SOMOS dataset and show the limitations that such models face when assigned to evaluate TTS utterances.

	\end{abstract}
	
	\section{Introduction}
	
	Recent advances in deep learning have resulted in the dominance of neural text-to-speech (TTS) systems in the field of speech synthesis, with state-of-the-art (SOTA) systems now approaching human-level quality. 
	However, synthetic speech evaluation still relies on subjective mean opinion score (MOS) tests, performed by human listeners who are entrusted with the task of rating speech utterances on a five-point absolute category scale.
	Despite being laborious and expensive, this procedure remains prevalent in TTS systems evaluation on speech quality and naturalness, since it cannot be substituted by the existing objective assessment metrics, which do not always correlate well with human perception.
	
	The development of objective metrics and models for the evaluation of generated speech relies heavily on the availability of large-scale subjective evaluation data and their respective synthetic stimuli. 
	Such data have been publicly released by the organizers of challenges that provide a common training set and a common listening test for researchers to compare their approaches.
	The Blizzard Challenge (BC) \cite{Black2005}, an annual workshop aiming to compare corpus-based speech synthesizers on the same data, has released data since 2008, which have been since used for the development of non-intrusive metrics. 
	The limited BC data from individual years have initially fostered efforts on the extraction of hand-crafted acoustic features which predict the quality of synthetic speech \cite{Falk2008, Hinterleitner2010, Norrenbrock2012}, while larger aggregated BC 2008-2013 data have been employed for feature extraction in the first attempt to use feed-forward networks for the MOS prediction task \cite{Yoshimura2016}.
	In the same year, AutoMOS \cite{Patton2016} explored deep recurrent architectures for predicting MOS values, which required a much larger proprietary dataset acquired over multiple years of TTS evaluation. 
	Quality-Net \cite{Fu2018}, based on bidirectional LSTM (BLSTM), was later trained on the TIMIT dataset to predict frame-level PESQ scores. 

	\subsection{Relevant Work}
	Nowadays, research focuses on training models for utterance-level and system-level evaluation predictions, in which feature extraction is automatically performed by the model itself, from the spectrograms provided as input to the model. 
	This paradigm requires adequately sized datasets.
	For TTS, the available BC data from early challenges (2008-2015) comprise systems of earlier TTS technologies, while data from latest challenges and modern synthesizers are multilingual \cite{Zhou2020, Ling2021}. 
	The aggregation of data from different tests is not expected to construct an inherently predictable dataset \cite{Hossfeld2016}, since scores are relative to contextual factors.
	
	The first large-scale public dataset of subjective evaluations from a single test was released by the Voice Conversion Challenge (VCC) in 2018 \cite{Lorenzo2018}, enabling the development of deep learning based models on open data, such as MOSNet \cite{Lo2019}, a CNN-BLSTM model for the evaluation of converted speech.
	MOSNet and the large VCC data have been since used as the basis for improving MOS prediction for both converted and synthesized speech \cite{Choi2020, Mittag2020, Leng2021}.
	However, \cite{Williams2020} report that pretrained VC models do not generalize well to TTS, and instead use data from ASVspoof 2019 \cite{Todisco2019}, which contain both synthetic and converted samples, for MOSNet training.
	Cooper \textit{et al}\cite{Cooper2021a} use the aforementioned pretrained model as an objective metric and find very weak correlation with human listener's evaluations.
	
	Responding to the need for standardized open datasets, the BVCC dataset \cite{Cooper2021a} was released as the basis of the VoiceMOS Challenge \cite{Huang2022}. 
	It comprises a wide variety of TTS systems, from unit selection to neural synthesis, as well as VC systems, evaluated in a single listening test by Japanese listeners.
	Due to the varied vocoding quality of this dataset, MOS naturalness scores are heavily influenced by the signal quality and artifacts, thus they cannot be interpreted as reflecting the samples' prosodic adequateness.
	BVCC consists of about 7K utterances, imposing constrains on its use for training parameter-heavy deep learning models, e.g. need for data augmentation techniques or fine-tuning of pretrained models.

	\subsection{Contribution}
	In this work, we describe the SOMOS dataset, a curated dataset of synthetic speech and subjective evaluations that can foster advancements in automatic MOS prediction research. 
	To the best of our knowledge, our work constitutes the first resource consisting of speech entirely produced by neural TTS models and their MOS ratings, adequately sized so as to train deep learning based systems that will be focused on acoustic model evaluation.
	Its scope is not to enable the training of a general purpose MOS prediction model, but rather assist research on TTS evaluation, in a single speaker scenario, where new TTS system ideas trained on LJ Speech can be quickly evaluated using a MOS prediction model developed from the SOMOS dataset.
	
	The remainder of this document is organized as follows. 
	In Section \ref{sec:speechdata}, we present the SOMOS speech dataset design. 
	Section \ref{sec:mosdata} elaborates on the construction of the subjective evaluations' dataset and the listening test results. 
	The performance of modern MOS prediction models trained on the SOMOS dataset is discussed in Section \ref{sec:baselines}, and Section \ref{sec:conclusion} concludes the paper indicating directions for future work.

	\section{Speech Dataset}
	\label{sec:speechdata}
	
	We design an English, single speaker dataset, as research has shown that listeners' preferences on a speaker's voice has significant effect on their perceived speech quality \cite{Hinterleitner2014}.
	Regarding the synthetic dataset’s voice, we have selected LJ Speech \cite{LJspeech17}, a public domain speech dataset with satisfactory quality, which is traditionally used for training benchmark systems in the TTS community.
	The speech dataset consists of 20,100 utterances of 201 systems (200 TTS systems described in Section~\ref{sec:systems} and the natural LJ Speech) and 2,000 unique sentences (Section~\ref{sec:sentences}).
	
	\subsection{Systems}
	\label{sec:systems}
	In order to synthesize speech utterances for our dataset, where the focus is on speech naturalness based on voice prosody rather than vocoding quality, we have used the same vocoder, LPCNet \cite{Valin2019, Vipperla2020} trained on LJ Speech, with all the TTS acoustic models.
	Our goal was not to include all existing acoustic models, but rather induce prosodic variability, thus we chose Tacotron-based models, which were familiar to us and suitable for producing multiple variations.
	The prosodically diverse synthetic utterances were generated by implicitly or explicitly varying the models' prosodic parameters, such as intonation, stress, rhythm and pauses.
	In total, we have included 200 different neural TTS ``systems".
	We define a ``system" as a neural TTS model with a fixed configuration of prosodic or style parameters.
	Thus, from a single TTS model that offers prosody or style manipulation capabilities, many ``systems" can be derived, by using a distinct configuration of parameters for each one.
	
	The neural acoustic models that are used to determine the 200 TTS systems are described below.
	A vanilla, attention-based architecture \cite{Ellinas2021} similar to Tacotron \cite{Wang2017, Shen2018} is used as a default neural TTS model, which typically produces output with a standard, average speaking style.
	In addition, a non-attentive Tacotron implementation based on \cite{Shen2020} is included.
	It is differentiated from its attention-based counterpart in rhythm and pauses, since phoneme durations are modelled and predicted with an explicit duration predictor.
	In an attempt to implicitly vary the prosody of our models through linguistic annotations, we have trained a Tacotron-like architecture with the addition of syntactic cues, where the input phoneme sequence has been enriched with noun chunk boundaries extracted from SpaCy's dependency parser \cite{Honnibal2015} and part of speech (POS) information from our frontend module. 
	These syntactic cues can also be interpreted as prosodic parameters that can be systematically altered to produce variation in speech intonation and pauses.
	Another approach to create additional prosodic variation, in a more explicit way, relies on ToBI (Tones and Break Indices) annotations \cite{Silverman1992} parallel to the input phoneme sequence.
	Our implementation is based on \cite{Zou2021}, but the neural ToBI prediction frontend is omitted.
	PyToBI \cite{Dominguez2019} is used to extract the original ToBI annotations, and by changing them in a systematic way, prosodically rich speech variants can be synthesized using this model.
	A non-attentive architecture with word-level, unsupervised style modelling \cite{Klapsas2021} has also been used, in order to induce prosodic changes strategically, in words, noun chunks and groups of chunks.
	The prosodic parameters chosen to be modified are stress, focus, energy, speed, coarticulation, and pauses.
	Finally, we have included a Tacotron-like model conditioned on five utterance-level prosodic speech features: pitch, pitch range, duration, energy, and spectral tilt \cite{Raitio2020}.
	Style variation is achieved by sampling the interpretable latent prosody space while focusing more on pitch range and spectral tilt.
	In this way, we have produced variations in sentence intonation, without overly distorting the inherent voice characteristics\footnote{Variant samples and info about the dataset's release can be found at: https://innoetics.github.io/publications/somos-dataset/index.html}.

	\subsection{Sentences}
	
	\begin{table}[!t]
		\footnotesize
		\caption{Domain, number and length in words (mean and standard deviation) of sentences used for inference.}
		\vspace{-5pt}
		\label{tab:sents}
		\centering
		\begin{tabular}[t]{@{\hspace*{1mm}} l @{\hspace*{2mm}} S @{\hspace*{0mm}} S @{\hspace*{0mm}} S @{\hspace*{1mm}}}
			
			\toprule
			\textbf{Domain} & \textbf{Count} & \multicolumn{2}{c}{\textbf{Length}}
			\\
			\midrule
			conv        & 306 & 10.9 & {\scriptsize${\pm}$5.3} \\
			general     & 119 & 11.8 & {\scriptsize${\pm}$1.8} \\
			Wikipedia   & 100 & 12.3 & {\scriptsize${\pm}$1.7} \\
			novel       & 362 &  7.4 & {\scriptsize${\pm}$1.7} \\		
			booksent    &  63 & 18.5 & {\scriptsize${\pm}$5.7} \\
			\bottomrule
		\end{tabular}
		\quad
		\begin{tabular}[t]{@{\hspace*{1mm}} l @{\hspace*{2mm}} S @{\hspace*{0mm}} S @{\hspace*{0mm}} S @{\hspace*{1mm}}}
			\toprule
			\textbf{Domain} & \textbf{Count} & \multicolumn{2}{c}{\textbf{Length}}
			\\
			\midrule
			reportorial & 425 & 21.6 & {\scriptsize${\pm}$7.3} \\
			news        & 391 & 10.1 & {\scriptsize${\pm}$2.8} \\
			broadcast   & 34 &  6.8 & {\scriptsize${\pm}$2.1} \\
			SUS         & 100 &  6.8 & {\scriptsize${\pm}$0.7} \\
			LJ Speech   & 100 &  17.0 & {\scriptsize${\pm}$5.9} \\
			\bottomrule
		\end{tabular}
		\vspace{-15pt}
	\end{table}
	
	\label{sec:sentences}
	We opted to create an inference corpus with a variety of linguistic contents for synthesis, so that the models that would be trained on the generated speech corpus would be able to generalize over it.
	The corpus comprises 2,000 sentences, out of which, 100 were randomly selected from the LJ Speech script and excluded from the LJ Speech training dataset. 
	The remainder of the corpus is composed in such a way to ensure domain, length and phoneme coverage.
	English sentences from the Blizzard Challenges 
	of years 2007-2016 were selected, mainly from the conversational, news, reportorial and novel domains, as well as 100 semantically unpredictable sentences (SUS).
	Additionally, Wikipedia and ``general'' public domain sentences from the web, which are common for conversational agents, have been included.
	The length of the corpus' sentences ranges from 3 to 38 words, with a median of 10 words.
	More details can be found in Table~\ref{tab:sents}.
	In our speech dataset, each system utters 100 sentences and each sentence is uttered by 10 distinct systems, except for the LJ Speech sentences which are uttered by 10 TTS systems and the natural LJ Speech voice.

	\section{Mean Opinion Scores' Dataset}
	\label{sec:mosdata}
	
	We have run a large-scale MOS test and gathered 395,318 crowdsourced ratings for our speech dataset, of which 359,380 concern the synthetic utterances. To our knowledge, this is the largest open subjective evaluations' dataset exclusively for neural TTS models. 
	
	\subsection{Listening Test Design}
	\label{sec:design}
	
	All subjective evaluations were conducted online with crowdsourced naive listeners via Amazon Mechanical Turk (AMT) \cite{Crowston2012}. 
	Listeners were asked to evaluate the naturalness of each sample on a 5-point Likert scale, while listening through headphones and being in a quiet setting. 
	The task was ``Rate how natural each sample sounds in a scale from 1 [very unnatural] to 5 [completely natural]" and the scale options were: ``1: very unnatural", ``2: somewhat unnatural", ``3: neither natural nor unnatural", ``4: somewhat natural", ``5: completely natural".
	To control for potential spurious participants, in each test page we included: (a) a ground truth sample from the LJ Speech recordings, (b) a validation sample, where listeners were instructed by the synthetic voice to select a given response from 1 to 5, and (c) a reCAPTCHA one-box tick, to deter malicious spam workers from participating. 
	Each test page, or Human Intelligence Task (HIT) in AMT terms, comprised 22 audio samples, of which 20 were synthetic of distinct systems and distinct sentences, 1 was ground truth, and 1 was validation. 
	All audio of the page had to be played at least once for the worker to be able to submit the page. 
	The test samples were structured in such a way that in 10 successive test pages all 200 systems were included. 
	With regards to the workers' payment, we have considered the required time per HIT completion and ensured compliance with the minimum hourly wage per locale. 
	
	Three tests were published simultaneously, each in a distinct English locale; United States (US), United Kingdom (GB) and Canada (CA). 
	This choice was made as we wanted our results to generalize to a varied population, as well as to allow for the identification of patterns in certain locales. 
	We also attempted to recruit Australian workers. However, we were unable to collect a sufficient number of scores from the AU locale.
	Only residents of the respective locale, according to AMT's qualification, were recruited in each test, with a past HIT approval rate of at least 90\%.
	We required only native English speakers to participate, but as this requirement could not be enforced, we have included a native/non-native checkbox in each test page.
	Every audio sample was evaluated by at least 17 unique participants (up to 23), of which at least 10 were US, 5 GB and 2 CA. In total, 987 workers participated in the evaluation. 
	The US test was completed in 4 hours by 785 workers; the GB test attracted 151 workers and took 3 weeks to complete; in CA, 51 listeners participated in the course of 2 weeks.

	\subsection{Quality of Ratings}
	\label{sec:quality}
	
\begin{figure}[!t]
		\captionsetup[subfigure]{labelformat=empty}
		\centering
		\subfloat[\centering]{\includegraphics[width=3cm]{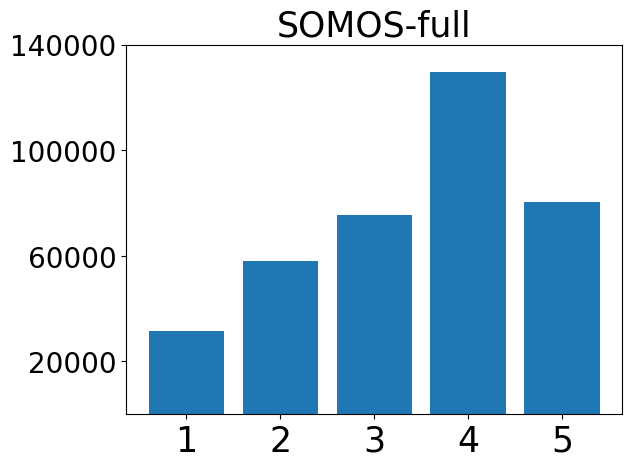}}%
		\qquad
		\subfloat[\centering]{\includegraphics[width=3cm]{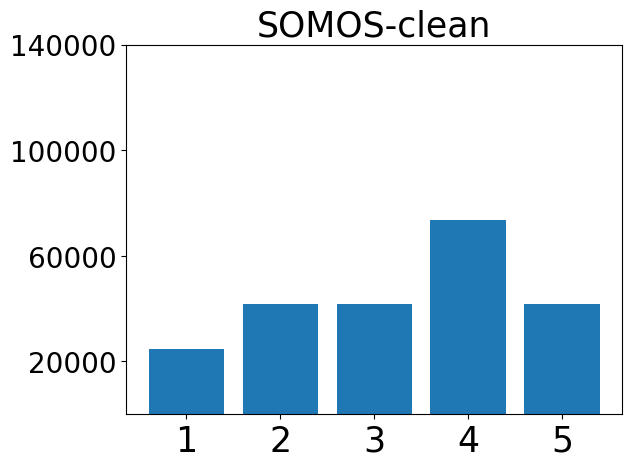}}%
		\vspace{-18pt}
		\caption{Distributions of scores for SOMOS-full and -clean}%
		\label{fig:distribution}
		\vspace*{-7mm}
	\end{figure}

	Although crowdsourcing is a quick way to collect a large number of annotations for language tasks, it has been shown that workers are not always reliable \cite{Snow2008}. 
	This is even more prominent for speech naturalness ratings, where the annotations consist of subjective opinions, and there is not one universally accepted answer.
	To mitigate this problem, we have followed the design recommendations in \cite{Kittur2008}, by inserting explicitly verifiable validation questions, and in \cite{Callison2010}, by inputting gold standard data (natural speech) in each test page.
	Additionally, by enabling page submission only if all audio on the page has been played, we have constrained users to spend on each page the same amount of time that would be required to complete the task honestly, which has been demonstrated to improve quality of crowdsourced responses \cite{Kittur2008}.
	Specifically for our task, through manual review of a small number of HITs, we have identified certain patterns that can be combined to indicate potentially unreliable responses:
	first, pages where all scores are the same, excluding that of the validation utterance, even more so since all samples have been generated from different systems as per the test's design;
	second, pages where the average synthetic score is higher or similar (smaller by 0.1) to the natural sample.
	These can be combined with the miss of the validation utterance score and the extremely low scores (1 or 2) given on natural utterances, in order to filter out suspicious HITs and create a ``clean" subset of the dataset. 
	The percentages of HITs per locale falling in any of the above \textit{quality control} categories are displayed in Table~\ref{tab:quality}.
	They do not add up to 100 as the controls are not mutually exclusive. 
	As can be observed, the US locale bears the larger portion of suspicious HITs, with half of the collected data being potentially unreliable. On the other hand, CA workers' consistency is remarkable, with almost no HITs falling into 2 out of 4 control categories.
	We hence refer to \textit{SOMOS-clean} to allude to the subset of the dataset that does not abide to any of the suspicious patterns, as opposed to \textit{SOMOS-full}, the dataset comprising all crowdsourced scores.
	The distributions of scores for the two are illustrated in Figure~\ref{fig:distribution}.
	\vspace{-2.5mm}
	\begin{table}[th]
		\footnotesize
		\caption{Quality controls on crowdsourced ratings per locale}
		\vspace{-5pt}
		\label{tab:quality}
		\centering
		\begin{tabular}{@{}lSSSS@{}}
			\toprule
			\multicolumn{1}{c}{} & \multicolumn{4}{c}{\textbf{Submitted HITs (\%)}}\\
			\midrule
			\textbf{Control}  & \textbf{US} & \textbf{GB} & \textbf{CA} & \textbf{All} \\
			\midrule
			Wrong validation  & 12.1 &  2.8 &  0.2 &  7.8  \\
			Low natural       & 10.8 & 10.4 &  2.4 &  9.7  \\
			Same scores       &  2.5 &  6.3 &  0.0 &  3.4  \\
			Similar/higher than natural    & 45.5 & 32.0 &  5.7 & 36.4  \\
			\midrule
			None of above      & 49.0 & 66.2 & 93.5 & 59.8  \\
			\bottomrule
		\end{tabular}
		\vspace{-8pt}
	\end{table}
	
	We have also collected the responses of an experienced US native linguist for $\approx$100 successive pages of the test, comprising all 200 TTS systems 10 times. 
	As expected, 100\% of their HITs do not fall in any suspicious category.
	Although the expert's scores average lower than crowdsourced scores, a large system-level positive relation is observed with all locales' results, using both the Pearson (PCC) and the Spearman rank correlation coefficient (SRCC). 
	This suggests that the test design has led to consistent answers that match the expert judgements.
	The locale with strongest correlation to expert ratings is GB (SRCC at \textit{r} = 0.88, PCC at \textit{r} = 0.84), while the weakest correlation is observed in US (SRCC at \textit{r} = 0.85, PCC at \textit{r} = 0.81).
	Interestingly, the correlation to the expert gets stronger for the entire dataset (SRCC at \textit{r} = 0.90) 
	suggesting the importance of combining ratings of several workers, as also showcased in \cite{Novotney2010, Marge2010}.

	\subsection{Results}
	\label{sec:results}
	
	\begin{figure*}[th]
		\centering
		\includegraphics[width=\textwidth]{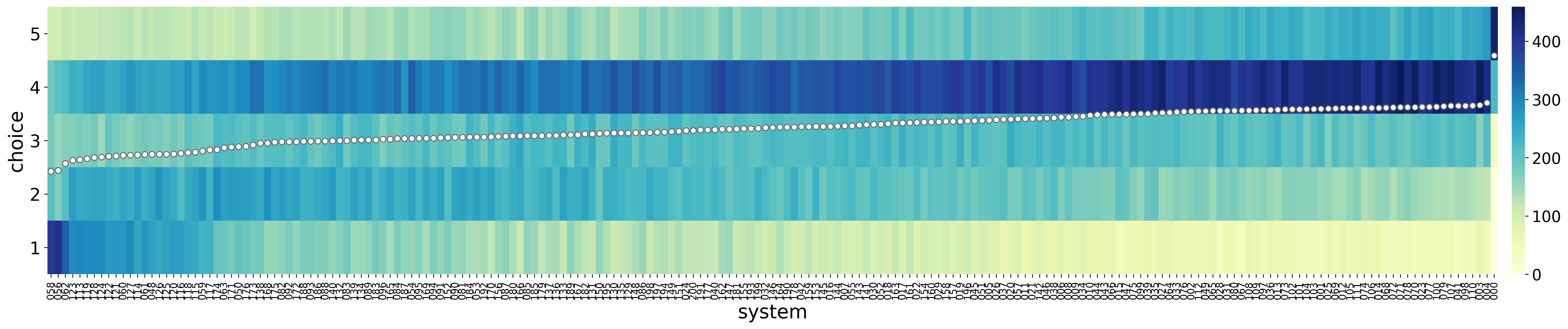}
		\vspace{-6mm}
		\caption{Raw MOS scores and average score per system for SOMOS-clean dataset}
		\label{fig:cleanresults}
		\vspace{-0.9em}
	\end{figure*}

	A histogram of the raw ratings for all 200 systems and natural speech in SOMOS-clean, arranged from lowest to highest average MOS score per system, can be seen in Figure~\ref{fig:cleanresults}.
	The darker colour concentration corresponds to more assembled ratings in the respective MOS choice for the respective system.
	The white dots depict the average MOS score per system.
	
	In order to determine the inherent correlation among listeners, we have used the bootstrap method \cite{Lo2019, Bisani2004}, in SOMOS-full and SOMOS-clean separately.
	For each of 1000 iterations, we randomly select and exclude half of the listeners, and calculate the linear correlation coefficient (LCC), Spearman's rank correlation coefficient (SRCC) and mean square error (MSE) on utterance-level and system-level average scores, between the resulting partial dataset and the whole dataset.
	The mean metrics over 1000 iterations are presented in Table~\ref{tab:inherentcorr}.
	
	\vspace{-2.5mm}
	\begin{table}[hbt!]
		\footnotesize
		\caption{Inherent correlations of human evaluations for SOMOS-full and SOMOS-clean at utterance and system level}
		\vspace{-2mm}
		\label{tab:inherentcorr}
		\centering
		\begin{tabular}{@{\hspace*{1mm}} l @{\hspace*{2mm}} r @{\hspace*{2mm}} r @{\hspace*{2mm}} r @{\hspace*{2mm}} r @{\hspace*{2mm}} r @{\hspace*{2mm}} r @{\hspace*{1mm}}}
			\toprule
			\multicolumn{1}{c}{} & \multicolumn{3}{c}{\textbf{utterance-level}} & \multicolumn{3}{c}{\textbf{system-level}} \\
			\midrule
			\textbf{}  & \textbf{MSE} & \textbf{LCC} & \textbf{SRCC} & \textbf{MSE} & \textbf{LCC} & \textbf{SRCC} \\
			\midrule
			SOMOS-full & \textbf{0.091} & 0.787 & 0.780 & \textbf{0.004} & 0.987 & 0.983 \\
			SOMOS-clean & 0.157 & \textbf{0.822} & \textbf{0.822} & 0.006 & \textbf{0.991} & \textbf{0.988} \\		
			\bottomrule
		\end{tabular}
		\vspace{-0.8em}
	\end{table}
	
	\vspace{-1mm}
	\section{MOS Prediction Baselines}	
	\label{sec:baselines}
	As there has been significant research on neural MOS prediction, we assess the performance of 3 SOTA models, namely MOSNet \cite{Lo2019}, LDNet \cite{Huang2021} and SSL-MOS \cite{Cooper2021b}, on our dataset.
	A preliminary split of SOMOS-full/clean into training (85\%) and validation (15\%) sets is used for these experiments.\footnote{With the dataset's release, we provide a carefully designed train-validation-test split (70\%-15\%-15\%) with unseen systems, listeners and texts, which can be used for further experimentation.} 
	In order to cover a variety of TTS systems, the validation set contains all utterances of 13 systems unseen with respect to the training set as well as utterances of 19 seen systems. 
	The MOS prediction models are trained from scratch on the preliminary split and their performance is evaluated by calculating the MSE, LCC and SRCC.
	Also, openly available checkpoints for each model pretrained in different datasets are evaluated on the SOMOS-clean validation set.
	Results can be viewed in Table~\ref{tab:baselines}.
	
	The clean version of the dataset yields better performance on all cases. 
	As expected, pretrained models on converted and synthetic speech of older technologies perform poorly on solely TTS data. 
	VC systems convert the speaker identity of a given utterance while maintaining its linguistic content and prosody, while older TTS systems present artifacts and discontinuities.
	As human evaluators presented with such samples may often attribute naturalness to vocoding quality, these data are not suitable for learning to predict the MOS of prosodically variant synthetic utterances.
	On the model side, SSL-MOS has the best overall performance, even on its pretrained version, confirming its superior generalization capability compared to other models.  
	
	The high system-level performances observed can be attributed to the evaluation setup, as the validation set contains systems which are seen during training and the unseen systems have similar model variants in the training set. 
	To confirm this hypothesis, we have experimented with subtracting an entire model's variants (41 systems) from training, leading to LDNet system-level performance of SRCC dropping to 0.73 and utterance-level performance to 0.31. 
	An important finding is that while system-level correlation scores are high, utterance-level scores are consistently very low, even when the systems are seen, suggesting that the plain frame-level acoustic information that these models process is not adequate to accurately evaluate a synthetic utterance.
	Such low scores present a limitation of current MOS prediction models which is crucial for speech synthesis.
	Utterance-level evaluations are most important for TTS, as the ``holy grail" of modern speech synthesis is the prosodic pertinence of a given synthetic utterance to its content, and current obstacles towards naturalness are prosodic inconsistencies, such as wrong stress or intonation patterns.

	\begin{table}[t]
		\footnotesize
		\caption{Utterance-level and system-level prediction results of baseline MOS prediction models on SOMOS validation set}
		\vspace{-2mm}
		\label{tab:baselines}
		\centering
		\begin{tabular}{@{\hspace*{0.5mm}} l @{\hspace*{0.8mm}} r @{\hspace*{0.9mm}} r @{\hspace*{0.9mm}} r @{\hspace*{1.3mm}} r @{\hspace*{0.6mm}} r @{\hspace*{0.8mm}} r @{\hspace*{0.4mm}}}
			\toprule
			\multicolumn{1}{c}{} & \multicolumn{3}{c}{\textbf{utterance-level}} & \multicolumn{3}{c}{\textbf{system-level}} \\
			\midrule
			\textbf{Model} (trainset) & \textbf{MSE} & \textbf{LCC} & \textbf{SRCC} & \textbf{MSE} & \textbf{LCC} & \textbf{SRCC} \\
			\midrule
			MOSNet (VCC2018) & 0.843 & -0.075 & -0.084 & 0.167 & -0.390 & -0.418 \\
			MOSNet (SOMOS-full) & \textbf{0.598} & 0.218 & 0.238 &\textbf{ 0.035} & 0.758 & 0.839 \\
			MOSNet (SOMOS-clean) & 0.729 & \textbf{0.352} & \textbf{0.347} & 0.123 & \textbf{0.815} & \textbf{0.816} \\
			\midrule
			LDNet  (BVCC) & 1.011 & 0.040 & 0.032 & 0.221 & 0.369 & 0.354 \\
			LDNet  (SOMOS-full) & \textbf{0.581} & 0.262 & 0.275 & \textbf{0.027} & 0.850 & 0.854 \\
			LDNet  (SOMOS-clean) & 0.642 & \textbf{0.397} & \textbf{0.386} & 0.034 & \textbf{0.950} & \textbf{0.905} \\
			\midrule
			SSL-MOS    (BVCC) & 2.217 & 0.229 & 0.230 & 1.597 & 0.792 & 0.805 \\
			SSL-MOS    (SOMOS-full) & \textbf{0.564} & 0.296 & 0.313 & \textbf{0.016} & 0.846 & 0.893 \\
			SSL-MOS    (SOMOS-clean) & 0.625 & \textbf{0.453} & \textbf{0.444} & 0.041 & \textbf{0.977} & \textbf{0.947} \\
			\bottomrule
		\end{tabular}
		\vspace{-7mm}
	\end{table}

	\vspace{-1mm}
	\section{Conclusions}
	\label{sec:conclusion}
	We have presented the SOMOS dataset, a dataset of TTS-synthesized speech and crowdsourced MOS naturalness evaluations.
	We have elaborated on the speech dataset's design, comprising 200 systems derived from Tacotron-based models suitable for prosodic variation and 2,000 sentences with domain coverage.
	We have gathered a large-scale dataset of MOS evaluations via crowdsourcing in 3 English AMT locales, and have reported in detail our test design as well as the measures taken to ensure the reliability of crowdsourced data.
	Finally, we have trained and tested state-of-the-art MOS prediction models on our dataset and have demonstrated the challenges they face in utterance-level evaluation as regards synthesized utterances with prosodic variability.
	In future work, our goal is to utilize this dataset so as to develop models which 
	correlate better to human evaluators on the utterance-level TTS assessment task.
	
	\vspace{-1mm}
	\section{Acknowledgements}
	We thank our colleagues Georgios Vardaxoglou at Innoetics, Patrick Hegarty and Srinivas Ponakala at Samsung Research America who provided insight and expertise of great assistance.

	\bibliographystyle{IEEEtran}
	
	\bibliography{mybib}

\end{document}